\documentclass[twocolumn,showkeys,showpacs,prb]{revtex4}
\usepackage{amsfonts, amsmath, latexsym}
\usepackage[dvips]{graphicx}

\newtheorem{1}{Proposition}
\newtheorem{2}{Theorem}
\newtheorem{3}{Corollary}
\newtheorem{4}[2]{Theorem}
\newtheorem{5}[2]{Theorem}
\newtheorem{6}[2]{Theorem}

\begin{document}

\title{The partial averaging method}

\author{Cristian Predescu}
\email{Cristian_Predescu@Brown.EDU} 
\affiliation{
Department of Chemistry, Brown University, Providence, Rhode Island 02912
}

\date{\today}
\begin{abstract}
The partial averaging technique is defined and used in conjunction with the random series implementation of the Feynman-Ka\c{c} formula. It enjoys certain properties such as good rates of convergence and convergence for potentials with coulombic singularities. In this work, I introduce the reader to the technique and I analyze the basic mathematical properties of the method. I show that the method is convergent for all Kato-class potentials that have finite Gaussian transform. 
\end{abstract}
\pacs{05.30.-d, 02.70.Ss}
\keywords{path integrals, partial averaging, random series, coulombic potentials}
\maketitle

\section{Introduction} 
\newcommand{\ud}{\mathrm{d}}
The thermodynamic properties of a monodimensional spinless quantum system characterized by the inverse temperature $\beta=1/(k_{B}T)$ are completely determined by the canonical partition function
\begin{equation}
\label{eq:1}
Z(\beta)=\int_{\mathbb{R}}{\rho(x,x;\beta)\ud x},
\end{equation}
where the (unnormalized) density matrix \[\rho(x,x';\beta)=\langle x|e^{-\beta H}|x'\rangle\] can be computed with the help of the Feynman-Ka\c{c} representation formula\cite{Fey48,Kac51} 
\begin{equation}
\label{eq:2}
\frac{\rho(x,x';\beta)}{\rho_{fp}(x,x';\beta)}=\mathbb{E}\exp\left\{-\beta\int_{0}^{1}\! \!  V\Big[x_0(u)+\sigma  B_u^0 \Big]\ud u\right\}
\end{equation}
for a large class of potentials. In Eq.~(\ref{eq:2}), $m_0$ is the mass of the particle, $x_0(u)$ is a shorthand for $x+(x'-x)u$, $\sigma= (\hbar^2\beta  /m_0)^{1/2}$, and \[\rho_{fp}(x,x';\beta)=\frac{1}{\sqrt{2\pi\sigma^2}} \exp{\left[-\frac{(x-x')^2} {2\sigma^2} \right]}\] is the density matrix of a similar free particle. 
The stochastic element that appears in Eq.~(\ref{eq:2}), $\{B_u^0,\, 0 \leq u \leq 1\}$, is a so-called standard Brownian bridge defined as follows: if 
$\{B_u,\, u\geq
0\}$ is a standard Brownian motion starting at zero, then the Brownian
bridge is the stochastic process~$\{B_u-u B_1 , 0 \leq u \leq 1\}$. Unless otherwise stated, in this paper, we shall reserve the symbol~$\mathbb{E}$ to denote the expected
value (average value) of  a certain random variable against the
underlying probability measure of the Brownian bridge~$B_u^0$. 

The generalization of the Eq.~(\ref{eq:2}) to a $d$-dimensional
system is straightforward. The symbol $B_u^0$ now denotes a
$d$-dimensional standard Brownian bridge, which is a vector $(B_{u,1}^0,
B_{u,2}^0,\ldots, B_{u, d}^0)$ with the components being independent
standard Brownian bridges. The symbol $\sigma$ stands for the vector
$(\sigma_1, \sigma_2, \ldots, \sigma_d)$ with components defined by
$\sigma_i^2=\hbar^2\beta/m_{0,i}$. The product $\sigma B_u^0$ is interpreted as the 
$d$-dimensional vector of components $\sigma_i B_{u,i}^0$.
Finally, $x$ and $x'$ are points in the configuration space
$\mathbb{R}^d$ connected by the line $x_0(u)=x+(x'-x)u$.

 As emphasized in Ref.~(\onlinecite{Dol99r}), the success of the Feynman-Ka\c{c} formula in the computation of the thermodynamic properties of quantum systems is fortuitously due to another remarkable result: the Metropolis \emph{et al.}\cite{Met53} sampling algorithm of arbitrary finite-dimensional probability distributions, which lies at the heart of the Monte Carlo integration schemes.\cite{Kal86} This leads to the related but separate problem of finding a rapidly convergent sequence of finite-dimensional approximations of the stochastic integral~(\ref{eq:2}). The main techniques found in literature can be classified in two categories: the discrete path integral methods and the random series techniques [for reviews, see Refs.~(\onlinecite{Mie01}) and (\onlinecite{Pre02}), respectively].  The latter methods lend themselves to various modifications which result in convergence for a wider class of potentials $V(x)$ or/and improved asymptotic convergence. One such method is the partial averaging (PA) technique, which was initially introduced by Doll, Coalson, and Freeman \cite{Dol85} as a way to accelerate the convergence of the ``primitive'' Fourier path integral method (FPI).\cite{Dol84} 
 
As we shall see, the partial averaging method requires  the Gaussian transform of the potential $V$ for its implementation. For real life potentials this is a difficult but not impossible task. However, it was generally considered that the improvement the technique brings in does not warrant the effort of computing the Gaussian transform of the potential and therefore,  the so-called gradient corrected partial averaging method was used instead. It has been shown that this latter method has general $\mathcal{O}(1/n^2)$ asymptotic behavior for sufficiently smooth potentials  and it has been argued that there is not much reason to suspect a better convergence rate for the full partial averaging method.\cite{Ele99} However, more accurate numerical evidence recently presented in Ref.~(\onlinecite{Pre02}) suggests that the full partial averaging method does have in fact a better behavior: if the technique is used in conjunction with the FPI approach and if the potential is  smooth enough, the asymptotic convergence  is $\mathcal{O}(1/n^3)$. The importance of the partial averaging method resides also in the fact that it acts as a prototypical strategy for improving the asymptotic rate of convergence of the random series path integral methods. As such, the reweighted random series technique\cite{Pre02} achieves superior asymptotic convergence  by simulating the  partial averaging approach. 

In this work, I shall argue for one more property of the partial averaging method which is not shared by the gradient corrected version and in general by the non-averaged methods. More specifically, I shall show that the method can be employed for potentials having negative coulombic singularities,  for which standard  discrete path integral techniques (and also the primitive random series ones) fail to converge.\cite{Law69, Kle95} In this respect, it is quite surprising that the technique has been scarcely used for this purpose, despite the fact that in several instances its application for the polaron problem\cite{Ale90, Tit01} and for the computation of statistical properties of quantum systems\cite{Kol01} was numerically successful. Though several other methods for dealing with the coulombic singularities have been proposed (see for instance Ref.~\onlinecite{Mus97}), I appreciate that the advantage of the partial averaging strategy can be best emphasized by its ability of handling such systems. 

In this paper,  I establish a sufficiently large class of potentials for which the partial averaging sequence of approximations converges to the correct Feynman-Ka\c{c} result, though I only study the convergence of the density matrix and of  the partition function. This class includes most of the smooth and bounded from below potentials as well as most of the potentials having coulombic singularities. The proofs I perform are of a rather trivial nature, as they are direct consequences of well established convergence theorems from general probability theory. In fact, they exploit the martingale property of the partial averaging method. Besides their intrinsic value, these convergence theorems are important because they set the proper mathematical context in which the partial averaging method should be further discussed or utilized.

\section{Formulation of the problem}
\subsection{A chemically relevant class of (scalar) potentials}
 A sufficiently large class of potentials for which the Feynman-Ka\c{c} formula (\ref{eq:2}) and its multidimensional analogues hold  is the so-called Kato class,  which we define below. If 
\[
g(y)=\left\{\begin{array}{cc} |y| &  d=1, \\ \ln({\|y\|^{-1}})& d=2, \\ \|y\|^{2-d}& d\geq 3, \end{array}\right.
\] 
then the Kato class $K_d$ is made up of all measurable functions $f: \mathbb{R}^d \to \mathbb{R}$ such that 
\begin{equation}
\label{eq:3}
\lim_{\alpha \downarrow 0} \sup_{x\in \mathbb{R}^d} \int_{\|x-y\|\leq \alpha}|f(y)g(x-y)|\ud y =0. 
\end{equation} 
We also say that $f$ is in $K_{d}^{\text{loc}}$ if $1_{D}f \in K_{d}$ for all bounded domains $D \subset \mathbb{R}^d$. We say that the potential $V(x)$ is of Kato class if its negative part $V_{-}=\max\{0, -V\}$ is in $K_d$, while its positive part $V_{+}=\max\{0,V\}$ is in $K_d^{\text{loc}}$.  In these conditions, as shown in Refs.~(\onlinecite{Sim79}) and (\onlinecite{Sim82}), the Feynman-Ka\c{c} formula~(\ref{eq:2}) holds. Moreover, Th.~B.7.5 of (Ref.~\onlinecite{Sim82}) shows that the density matrix $\rho(x,x';\beta)$ is continuous on  $ \mathbb{R}^{d}\times \mathbb{R}^{d}\times (0, \infty)$, while Th.~B.6.7 of the same reference shows that for a given $\beta>0$, the density matrix is uniformly bounded in the variables $(x,x')$.

A remarkable theorem due to Aizenman and Simon,\cite{Aiz82} gives an alternative definition for the Kato class $K_d$. More precisely, Theorem~4.5 of Ref.~(\onlinecite{Aiz82}) says that $V\in K_d$ if and only if 
\begin{equation}
\label{eq:3a}
\lim_{\epsilon \downarrow 0} \sup_{x\in \mathbb{R}^d} \mathbb{E}\left[ \int_0^{\epsilon}|V(x+\sigma B_u)|\ud u\right] =0,
\end{equation}
where  $\mathbb{E}$ denotes the expectation value against the  $d$-dimensional Brownian motion $B_u$. Inverting the order of integration in Eq.~(\ref{eq:3a}) and remembering that $B_u$ is a Gaussian distributed variable of variance $u$, we obtain the equivalent condition 
\begin{equation}
\label{eq:3c}
\lim_{\epsilon \downarrow 0} \sup_{x\in \mathbb{R}^d}   \int_0^{\epsilon}\ud u \int_{\mathbb{R}^d}{(2\pi u)^{-d/2}}e^{-\|z\|^2/(2u)}|V(x+\sigma  z)|\ud z =0.
\end{equation}
We leave it for the reader to perform  the substitutions $u'=u/\epsilon$ and then $z'= z/\sqrt{\epsilon}$ in successive order and prove the following reformulation of the condition given by Eq.~(\ref{eq:3c}):
\begin{equation}
\label{eq:3b}
\lim_{\epsilon \downarrow 0} \sup_{x\in \mathbb{R}^d}  \epsilon \int_0^{1}\ud u \int_{\mathbb{R}^d}{(2\pi u)^{-d/2}}e^{-\|z\|^2/(2u)}\left|V(x+\sigma \sqrt{\epsilon}  z)\right|\ud z =0.
\end{equation}
In the Appendix, we shall use Eq.~(\ref{eq:3a}) in the proof of Th.~\ref{th:3A} and Eq.~(\ref{eq:3b}) in the proof of Th.~\ref{th:2A}, respectively. 

As far as the chemical physicist is concerned, the Kato class of potentials is sufficiently general. It contains for instance the coulombic potential as it appears in electronic structure calculations. For another example, the ab initio intermolecular potential computed at the level of the Born-Oppenheimer approximation cannot have singularities worse than the coulombic ones and therefore it is of Kato class. However, we do not consider certain empirical potentials which are not of Kato class, as for example the Leonard-Jones potential. Nevertheless, this can be brought into the Kato class if the unphysical $r^{-12}$ singularity is removed by  truncation or by other approximations.  

Let us anticipate a little and also demand that the potential $V$ have finite Gaussian transform. The reader may read ahead in the next subsection and see that this condition is natural for the proper definition of the partial averaging method. More precisely, we require that 
\begin{eqnarray}
\label{eq:4}
\overline{|V|}_{\alpha}(x)=\left(\prod_{i=1}^d \frac{1}{{2\pi \alpha_{i}^2}}\right)^{1/2}\int_{\mathbb{R}}\ud z_1 \cdots  \int_{\mathbb{R}}\ud z_d \nonumber \\ \times \exp\left(-\sum_{i=1}^d \frac{z_i^2}{2\alpha_{i}^2}\right) \left|V(x+z)\right|  < \infty,
\end{eqnarray} 
for all $x\in \mathbb{R}^d$ and $\alpha \in \mathbb{R}_+^d$. [In this paper, $\mathbb{R}_+=(0,\infty)$.]
Certain properties of the potentials having finite Gaussian transform are given by Theorem~\ref{th:1A} of the Appendix.

From the thermodynamic point of view, only the diagonal density matrix  $\rho(x,x;\beta)$ is of interest. Moreover, in order to have a physically relevant statistics, the condition
\begin{equation}
\label{eq:5}
0< Z(\beta)<\infty; \quad \forall \beta>0
\end{equation}
must hold, but the inequality (\ref{eq:5}) is not a requirement for the results obtained in this paper to be valid. In practice, the condition is achieved by the addition of a \emph{constraining} potential, which is usually a continuous and bounded from below  function (thus still in the Kato class). The constraining potential is intended to simulate, for example,  the effect of the container in which a reaction takes place. A sufficient condition for the quantum partition function to be finite is that the analog classical partition function be finite. This follows from the following inequality:  
\begin{1}
\label{pr:1}
Set \begin{equation*}
Z_{cl}(\beta)=\frac{1}{\sqrt{2\pi \sigma^2}}\int_{\mathbb{R}}e^{-\beta V(x)}\ud x. 
\end{equation*}
Then,
$Z(\beta)\leq Z_{cl}(\beta)$.
\end{1}
\emph{Proof.} By Jensen's inequality and Tonelli theorem,
\begin{eqnarray*}
Z(\beta)= \frac{1}{\sqrt{2\pi \sigma^2}} \int_{\mathbb{R}}\ud x\, \mathbb{E}\,e^{-\beta \! \int_0^1 V(x+\sigma B_u^0)\ud u }  \\ \leq \frac{1}{\sqrt{2\pi \sigma^2}}\int_{\mathbb{R}}\ud x\, \mathbb{E} \!\int_0^1\! \ud u \, e^{-\beta V(x+\sigma B_u^0)} \\= \frac{1}{\sqrt{2\pi \sigma^2}}\mathbb{E} \int_0^1\! \ud u \! \int_{\mathbb{R}}\ud x\,  e^{-\beta V(x+\sigma B_u^0)}=Z_{cl}(\beta).\quad  \Box
\end{eqnarray*}

As stated, Proposition~(\ref{pr:1}) remains true  for multidimensional systems. In this paper, we shall perform the proofs only for monodimensional systems. The reader should notice that our arguments are purely measure theoretic, in fact irrespective of the dimensionality of the physical systems.  On the other hand, in the chemical physics literature it is customary to perform the analysis in ``monodimensional'' notation. I consider that the mathematician will have little trouble generalizing the proofs, yet the chemist may find it hard to accommodate a more complicated notation.  

\subsection{The partial averaging strategy}
	
	In this section, we shall give a short review of the partial averaging method for monodimensional systems (the multidimensional generalization is straightforward). For a more complete discussion, the reader should consult Ref.~(\onlinecite{Pre02}). The most general series representation of the Brownian bridge is given 
by the Ito-Nisio theorem,\cite{Kwa92} the statement of which is reproduced below. 
Assume given $\{\lambda_k(\tau)\}_{k \geq 1}$
a system of functions on the interval $[0,1]$ which, together with the constant function
 $\lambda_0(\tau)=1$, makes up an orthonormal basis in $L^2[0,1]$. Let $\Lambda_k(u)$ denote the primitives 
\[ \Lambda_k(u)=\int_0^u \lambda_k(\tau) \ud \tau\]
of the functions $\lambda_k(u)$. If $\Omega$ 
is the space of infinite sequences $\bar{a}\equiv(a_1,a_2,\ldots)$ and 
\begin{equation}
\label{eq:6}
P[\bar{a}]=\prod_{k=1}^{\infty}\mu(a_k)
\end{equation}
 is the (unique by the Kolmogorov extension theorem) probability measure on $\Omega$ such that the coordinate maps $\bar{a}\rightarrow a_k$ are independent identically distributed (i.i.d.) variables with distribution probability
\begin{equation}
\label{eq:7}
\mu(a_k\in A)= \frac{1}{\sqrt{2\pi}}\int_A e^{-z^2/2}\,\ud z,
\end{equation}
then
\begin{equation}
\label{eq:8}
B_u^0(\bar{a}) = \sum_{k=1}^{\infty}a_k\Lambda_{k}(u),\; 0\leq u\leq1
\end{equation}
is equal in distribution to a standard Brownian bridge.  Moreover, the convergence of the above series is almost surely uniform on the interval $[0,1]$.

Using the Ito-Nisio representation of the Brownian bridge, 
the Feynman-Ka\c{c} formula (\ref{eq:2}) takes the form
\begin{eqnarray}
\label{eq:9}
 \frac{\rho(x, x' ;\beta)}{\rho_{fp}(x, x' ;\beta)}&=&\int_{\Omega}\ud P[\bar{a}]\nonumber  \exp\bigg\{-\beta \int_{0}^{1}\! \!  V\Big[x_0(u) \\& +& \sigma \sum_{k=1}^{\infty}a_k \Lambda_k(u) \Big]\ud u\bigg\}.
\end{eqnarray}
The independence of the coordinates $a_k$, which physically amounts to choosing those representations in which the kinetic energy operator is diagonal, is the key to the use of the partial averaging method. Denoting by $\mathbb{E}_{n}$ the average over the coefficients beyond the rank~$n$, the partial averaging formula reads
\begin{eqnarray}
\label{eq:10}
 &&\frac{\rho_n^{{PA}}(x, x' ;\beta)}{\rho_{fp}(x, x' ;\beta)}=\int_{\mathbb{R}}\ud \mu(a_1)\ldots \int_{\mathbb{R}}\ud \mu(a_n)\nonumber \\ &&\times \exp\bigg\{-\beta \; \mathbb{E}_n\int_{0}^{1}\! \! 
V\Big[x_0(u)+\sigma \sum_{k=1}^{\infty}a_k
\Lambda_k(u) \Big]\ud u\bigg\}. \qquad
\end{eqnarray}
Assuming that the  Fubini theorem holds (this is proved in the next section), one may invert the order of integration in the exponent and compute
\begin{eqnarray}
\label{eq:11}&&
\mathbb{E}_n\!\int_0^1\!  V[x_0(u)+\sigma B_u^0(\bar{a})]\ud u\nonumber \\&& = \int_0^1\!  \mathbb{E}_n V[x_0(u)+\sigma B_u^0(\bar{a})] \ud u \\&&=\int_0^1\!  \overline{V}_{u,n}[x_0(u)+\sigma \sum_{k=1}^n a_k \Lambda_k(u)]\ud u, \nonumber
\end{eqnarray}
where
\begin{equation}
\label{eq:12}
\overline{V}_{u,n}(y)=\int_{\mathbb{R}}\frac{1}{\sqrt{2\pi\Gamma_{n}^2(u)}} \exp\left[-\frac{z^2}{2\Gamma_{n}^2(u)}\right]V(y+z) \ud z.
\end{equation}
The function $\Gamma_n^2(u)$ is defined by
\begin{eqnarray}
\label{eq:13}
\Gamma_{n}^2(u)&=&\sigma^2
\sum_{k={n+1}}^{\infty}\Lambda_k(u)^2\nonumber \\ &=&\sigma^2\left[u(1-u)-
\sum_{k=1}^{n}\Lambda_k(u)^2\right].
\end{eqnarray}
[Again, the reader is refered to Ref.~(\onlinecite{Pre02}) for additional explanations. It is customary to use the notation $\Gamma_n^2(u)$ to mean the square of $\Gamma_n(u)$.] To summarize, we \emph{define} the $n$-th order partial averaging approximation to the diagonal density matrix by the formula
\begin{eqnarray}
\label{eq:14}&&
\frac{\rho^{PA}_n(x, x' ;\beta)}{\rho_{fp}(x, x' ;\beta)}=\int_{\mathbb{R}}\ud \mu(a_1)\ldots \int_{\mathbb{R}}\ud \mu(a_n)\nonumber  \\&& \times \exp\bigg\{-\beta \; \int_{0}^{1}\! \! 
\overline{V}_{u,n}\Big[x_0(u)+\sigma
\sum_{k=1}^{n}a_k \Lambda_k(u) \Big]\ud u\bigg\}.\qquad
\end{eqnarray}

\section{Martingale property and convergence results}

This section establishes the martingale property of the partial averaging method. One may notice that we obtain some important convergence results without actually doing much work other than citing some well-established theorems. The chemical physicist will probably  be more interested in the  Corollary~1, which is for that matter presented separately.
  	
On the set $\Omega$ of sequences $\bar{a}\equiv (a_1, a_2, \ldots)$, consider the $\sigma$-algebra generated by the finite-dimensional Borel sets $\mathcal{F}_{\infty}=\sigma(\cup_{n \geq 0}\mathcal{F}_n)$, where $\mathcal{F}_n=\sigma(a_1, a_2,\ldots,a_n)$ and $\mathcal{F}_0=\{\oslash,\Omega\}$. By construction, $\{\mathcal{F}_n\}_{n \geq 0}$ is a filtration. Also, the probability measure $\ud P[\bar{a}]$ introduced in the previous section is, of course, defined over $\mathcal{F}_{\infty}$ so that the default probability space we refer to in this work is $(\Omega, \mathcal{F}_{\infty}, P)$. If $f$ is an integrable random variable on~$\Omega$, we shall sometimes denote the conditional expectation $\mathbb{E}\, [f\,|\mathcal{F}_n]$ simply by~$\mathbb{E}_n f$.

To continue with the introduction of the notations, we define
\[
U_n(x,x',\beta; \bar{a})=\int_0^1\overline{V}_{u,n}\left[x_0(u)+\sigma \sum_{k=1}^n a_k \Lambda_k(u)\right] \ud u
\]
and
\[
U_\infty(x,x',\beta; \bar{a})=\int_0^1V\left[x_0(u)+\sigma\sum_{k=1}^\infty a_k \Lambda_k(u) \right] \ud u.
\]
The variables $x$, $x'$, and $\beta$ are interpreted here as parameters and, just as a reminder, we separate them by a semicolon from the ``true'' variable $\bar{a}$. By construction, $U_n(x,x',\beta;\bar{a})$ is $\mathcal{F}_n$ measurable, while $U_\infty(x,x',\beta;\bar{a})$ is $\mathcal{F}_\infty$ measurable. 
Let us prove that $U_n(x,x',\beta;\bar{a})=\mathbb{E}_n \left[U_{\infty}(x,x',\beta;\bar{a})\right]$ for $P$-almost every $\bar{a}$. As shown in the previous section, this boils down to proving 
\begin{eqnarray*}
\mathbb{E}_n\!\int_0^1\!  V[x_0(u)+\sigma B_u^0(\bar{a})]\ud u\nonumber \\= \int_0^1 \! \mathbb{E}_n V[x_0(u)+\sigma B_u^0(\bar{a})]\ud u\nonumber
\end{eqnarray*}
$P$-almost surely. 
The relation follows from the Fubini-Tonelli theorem provided that 
\begin{equation}
\label{eq:15}
 \int_0^1 \!  \mathbb{E}_n \left|V[x_0(u)+\sigma B_u^0(\bar{a})]\right|\ud u 
\end{equation}
is finite $P$-a.s. Since the above integrand is nonnegative, it is enough to show that its $P$-expectation is finite. Let
\[\ud \mu_\epsilon(z)=\frac{1}{\sqrt{2\pi\epsilon^2}}e^{-z^2/(2\epsilon^2)}\ud z\] denote the respective Gaussian measure on $\mathbb{R}$. Using the Tonelli theorem to invert the order of integration, one computes
\begin{eqnarray}
\label{eq:16}&&
\nonumber  \mathbb{E}\left\{ \int_0^1 \!  \mathbb{E}_n \left|V[x_0(u)+\sigma B_u^0(\bar{a})]\right|\ud u\right\}\\&=& \int_0^1 \!  \mathbb{E}\left|V[x_0(u)+\sigma B_u^0(\bar{a})]\right|\ud u \\&=& \int_0^1 \!\ud u  \int_{\mathbb{R}}  \left|V[x_0(u)+z ]\right| \ud \mu_{\Gamma_0(u)}(z)< \infty.\nonumber
\end{eqnarray}
The last expression is finite for all $(x,x';\beta)\in \mathbb{R}^2\times \mathbb{R}_+$  by  Th.~\ref{th:2A} of the  Appendix.  Eq.~(\ref{eq:16}) also shows that $U_{\infty}(x,x',\beta;\bar{a})$ is $P$-integrable for all $(x,x';\beta)$. Then, standard theorems from martingale theory show [see Theorem~5.7 of Ref.~(\onlinecite{Dur96})]:
\begin{2}
\label{th:1}
For all $(x,x',\beta)\in \mathbb{R}^2\times \mathbb{R}_+$, the sequence $U_{n}(x,x',\beta;\bar{a})$ is a martingale adapted to the filtration $\mathcal{F}_n$ and is a.s. and $L^1$ convergent to $\mathbb{E}\left[U_{\infty}(x,x',\beta;\bar{a})|\mathcal{F}_\infty\right]=U_{\infty}(x,x',\beta;\bar{a})$.
\end{2}

Let us define
\[X_n(x,x',\beta;\bar{a})=\rho_{fp}(x,x';\beta)\exp[-\beta \, U_n(x,x',\beta;\bar{a})]\]
and
\[X_\infty(x,x',\beta;\bar{a})=\rho_{fp}(x,x';\beta)\exp[-\beta\, U_\infty(x,x',\beta;\bar{a})].\]
Then, we have
\begin{equation}
 \label{eq:17} \rho(x,x';\beta)=\mathbb{E}\,[X_\infty(x,x',\beta;\bar{a})] 
\end{equation}
and
\begin{equation}
\label{eq:18} \rho_n^{PA}(x,x';\beta)=\mathbb{E}\,[X_n(x,x',\beta;\bar{a})],
\end{equation}
respectively.

\begin{2}[PA Convergence Theorem]
\label{th:2}
For all $(x,x',\beta)\in \mathbb{R}^2\times \mathbb{R}_+$, the sequence~$X_n(x,x',\beta;\bar{a})$  is a  submartingale adapted to the filtration~$\mathcal{F}_n$ and is a.s. and $L^1$ convergent to $X_\infty(x,x',\beta;\bar{a})$. 
\end{2}

\emph{Proof.} We notice that $X_n(x,x',\beta;\bar{a})$ is the exponential of a martingale. Thus, by the conditional Jensen's inequality [see page 225 of Ref.~(\onlinecite{Dur96})], we have 
\begin{equation}
\label{eq:fi}
X_n \leq \mathbb{E}\,[X_{n+1}|\mathcal{F}_n] \leq \mathbb{E}\,[X_{\infty}|\mathcal{F}_n].
\end{equation}
The above inequality establishes the submartingale property because, as mentioned in Section II.A,
\[\rho(x,x';\beta)=\mathbb{E}\,\big[\mathbb{E}\,[X_{\infty}|\mathcal{F}_n]\big]=\mathbb{E}X_{\infty}\]
is uniformly bounded in the variables $(x,x')$ for all $\beta >0$. An elementary proof of this assertion is given in the Appendix (see Th.~\ref{th:3A}). Finally, the a.s. convergence follows directly from Th.~\ref{th:1}, while the $L^1$ convergence follows from the Dominated convergence theorem and the inequality (\ref{eq:fi}). $\ \Box$

 We define the $n$-th order partial averaging partition function by the formula
\[
Z_n^{PA}(\beta)=\int_{\mathbb{R}} \rho_n^{PA}(x,x;\beta)\ud x.
\] 
Using the symbol $\uparrow$ to mean ``monotonically increasing to,'' a direct consequence of Th.~\ref{th:2} is the following. 
\begin{3}
\label{co:1}
As $n \rightarrow \infty$,
\begin{equation}
\label{eq:19}
\rho_n^{PA}(x,x';\beta) \uparrow \rho(x,x';\beta)\ \text{and} \ Z_n^{PA}(\beta) \uparrow Z(\beta).
\end{equation}
\end{3}
\emph{Proof.}
The pointwise monotonic convergence of the density matrix is a direct consequence of the submartingale property and of the $L^1(\Omega,P)$ convergence of the partial averaging method. Then, the convergence of the partition functions follows from the Monotone convergence theorem.
$\ \Box$

\section{Summary}
In this paper, I presented the basic properties of the partial averaging method. I demonstrated that the method can be employed for a quite general class of potentials by proving several convergence results of interest for the chemical physicist. In particular, I proved that the PA method is convergent for most of the potentials having negative coulombic singularities. The value of the convergence theorems deduced in the present paper consists of the fact that they establish the mathematical context in which the partial averaging method should be utilized or discussed. I also anticipate that the martingale property will  play an important role in  establishing the asymptotic rates of convergence for different partial averaging schemes and, in fact, it may explain the superior asymptotic behavior of these methods.

 \begin{acknowledgments}
   The author acknowledges support from the National Science Foundation through 
awards CHE-0095053 and CHE-0131114.  He also would like to
thank Professor J. D. Doll for continuing discussions concerning the present developments.
\end{acknowledgments}

\appendix
\section{}

The following theorem consists of well known facts about the Gaussian transform and I present it here for ease of reference. 
\begin{4}
\label{th:1A}
Let $f:\mathbb{R}^d \rightarrow \mathbb{R}$ be a Borel measurable function, let $\alpha=(\alpha_0,\ldots,\alpha_d)\in \mathbb{R}_+^d$, and let $\|\alpha\|=\max_{1\leq i\leq d}\alpha_i $. Consider the $d$-dimensional Gaussian measure
\begin{equation}
\label{eq:1A}
\ud \mu_{\alpha}(z)=\prod_{i=1}^d\left[\frac{1}{\sqrt{2 \pi \alpha_i^2}}e^{-z_i^2/(2\alpha_i^2)}\ud z_i\right]
\end{equation}
and let
\[
F(x,\alpha)=\int_{\mathbb{R}^d} |f(x+z)|\ud \mu_\alpha(z).
\]
be defined on $\mathbb{R}^d\times \mathbb{R}_+^d$. Assume there is $(y, \eta) \in \mathbb{R}^d\times \mathbb{R}_+^d$ such that $F(y, \eta) < \infty$ and let $D= \mathbb{R}^d\times \prod_{i=1}^d (0,\eta_i)$. Then the following are true:
\begin{enumerate}
\item[a)]{f is locally integrable.}
\item[b)]{$F(x,\alpha)< \infty \quad  \text{for all} \; (x,\alpha)\in D$.}
\item[c)]{\[G(x,\alpha)=\int_{\mathbb{R}^d} f(x+z)\ud \mu_\alpha (z)\] is well defined, continuous and infinitely differentiable in both arguments on $D$.}
\item[d)]{$\lim_{\|\alpha\| \rightarrow 0} G(x,\alpha) = f(x) \text{ a.e. }$ More strongly, 
\[
\lim_{\|\alpha\| \rightarrow 0} \int_{\mathbb{R}^d} |f(x+z)-f(x)|\ud \mu_\alpha (z) =0  \text{ a.e. }
\]}
\end{enumerate} 
\end{4}

\begin{5}
\label{th:2A}
Let $x$ and $x'$ be arbitrary points in $\mathbb{R}^d$, let $\beta>0$, and let $B_u^0$ be a standard $d$-dimensional Brownian bridge on $[0,1]$. Pick some arbitrary $\sigma=(\sigma_1,\ldots,\sigma_d)\in \mathbb{R}_+^d$ and  set $y=x'-x$. If $V\in K^{loc}_d$ and $V$ has finite Gaussian transform, then 
\begin{eqnarray}
\label{eq:2A}&&
 \int_0^1 \!  \nonumber \mathbb{E}\left|V(x+yu+\sigma  B_u^0)\right|\ud u \\&=& \int_0^1 \!\ud u  \int_{\mathbb{R}^d}  \left|V(x+yu+z )\right| \ud \mu_{\Gamma_0(u)}(z)< \infty,\qquad
\end{eqnarray}
where $\Gamma^2_0(u)=u(1-u)(\sigma^2_1,\ldots,\sigma^2_d)$ and the Gaussian measure $\ud \mu_\alpha(z)$ is defined by the relation (\ref{eq:1A}). 
\end{5}

\emph{Observation.} We stated this theorem separately because its proof depends upon the dimensionality of the problem. More precisely, if the system is monodimensional, one may use $\Gamma_0^2(u)\leq \sigma^2$ to show that the integral (\ref{eq:2A}) is smaller than
\begin{equation}
\label{eq:3A}
\int_0^1 \!\ud u \frac{1}{\sqrt{u(1-u)}}\int_{\mathbb{R}}  \left|V(x+yu+z )\right| \ud \mu_{\sigma}(z).
\end{equation}
 By Theorem~\ref{th:1A}.c), the integral \[\int_{\mathbb{R}}  \left|V(x+yu+z )\right| \ud \mu_{\sigma}(z)\] as a function of $u$ is continuous on $[0,1]$, thus bounded.  Then, by the integrability of $[u(1-u)]^{-1/2}$, it follows that the integral (\ref{eq:3A}) is finite. However, this reasoning is not valid for higher dimensions because $[u(1-u)]^{-d/2}$ is not integrable for $d\geq 2$ and the additional condition $V \in K^{loc}_d$ is needed. 
 
\emph{Proof of the theorem.} The equality (\ref{eq:2A}) was discussed in the text. From Th.~\ref{th:1A}.c), it follows that 
\[
\int_{\mathbb{R}^d}  \left|V(x+yu+z )\right| \ud \mu_{\Gamma_0(u)}(z)
\]
as a function of $u$ is continuous on all compact intervals $[\epsilon, 1-\epsilon]$ with $0<\epsilon<1/2$, thus bounded and integrable. It is then enough to show that 
\begin{eqnarray*}
I_\epsilon(x,y)&=&\int_0^\epsilon \!\ud u  \int_{\mathbb{R}^d}  \left|V(x+yu+z )\right| \ud \mu_{\Gamma_0(u)}(z)\\&=&\int_0^\epsilon \!\ud u  \int_{\mathbb{R}^d}  \left|V(x+z )\right| \ud \mu_{\Gamma_0(u)}(z-yu)< \infty
\end{eqnarray*}
for all $x$ and $y$ and small enough $\epsilon$. This is so because the integral over the  end $[1-\epsilon, 1]$ can be shown to  equal  $I_\epsilon(x+y,-y)$ by the change of variable $u'=1-u$. We shall prove the above inequality in two steps.

\emph{Step~1.} In the first step, we prove the inequality
\begin{eqnarray}
\label{eq:4A}\nonumber
I_\epsilon(x,y)&\leq& 2^{d}\exp\left( \sum_{i=1}^d \frac{y_i^2}{4\sigma_i^2}\right)\\& \times& \int_0^\epsilon \!\ud u  \int_{\mathbb{R}^d}  \left|V(x+z)\right| \ud  \mu_{\sigma \sqrt{2u}}(z).
\end{eqnarray}
The inequality $0<u< 1/2$ implies
\begin{eqnarray}
\label{eq:5A}\nonumber
\prod_{i=1}^d\left\{\frac{1}{\sqrt{2 \pi \sigma_i^2 u(1-u)}}\exp\left[-\frac{(z_i-y_iu)^2}{2\sigma_i^2 u(1-u)}\right]\right\} \leq \\
2^{d/2}\prod_{i=1}^d\left\{\frac{1}{\sqrt{2 \pi \sigma_i^2 u}}\exp\left[-\frac{(z_i-y_iu)^2}{2\sigma_i^2 u}\right]\right\}. 
\end{eqnarray}
On the other hand, the minimum of the expression
\[
z_i^2/2-2z_iy_iu+y_i^2u^2
\]
as a quadratic function of $z_i$ is attained at $z_i=2y_iu$ and has the value $-y_i^2 u^2$. Therefore,
\begin{eqnarray*}\frac{(z_i-y_iu)^2}{2\sigma_i^2 u}=\frac{z_i^2/2+z_i^2/2-2z_iy_iu+y_i^2u^2}{2\sigma_i^2 u}\\ \geq \frac{z_i^2}{4\sigma_i^2 u}-\frac{y_i^2u}{2\sigma_i^2}\geq \frac{z_i^2}{4\sigma_i^2 u}-\frac{y_i^2}{4\sigma_i^2},  \end{eqnarray*}
where we used again the condition $u<1/2$. 
Replacing the last inequality in Eq.~(\ref{eq:5A}), we obtain
\begin{eqnarray*}
\prod_{i=1}^d\left\{\frac{1}{\sqrt{2 \pi \sigma_i^2 u(1-u)}}\exp\left[-\frac{(z_i-y_iu)^2}{2\sigma_i^2 u(1-u)}\right]\right\} \leq \\
2^{d} \exp\left( \sum_{i=1}^d \frac{y_i^2}{4\sigma_i^2}\right)\prod_{i=1}^d\left[\frac{1}{\sqrt{4 \pi \sigma_i^2 u}}\exp\left(-\frac{z_i^2}{4\sigma_i^2 u}\right)\right] 
\end{eqnarray*}
and, consequently, the inequality given by Eq.~(\ref{eq:4A}) is proven. 

\emph{Step~2.}
By the results from the first step, it suffices to show that the last integral in Eq.~(\ref{eq:4A}) is finite. By appropriate transformation of coordinates, we may rewrite this last integral as
\begin{eqnarray*}&&
\int_0^\epsilon \!\ud u  \int_{\mathbb{R}^d}  \left|V(x+z)\right| \ud  \mu_{\sigma \sqrt{2u}}(z) \\ &&= \epsilon \int_0^1 \!\ud u  \int_{\mathbb{R}^d}  \left|V(x+z)\right| \ud  \mu_{\sigma \sqrt{2\epsilon u}}(z) \\&&=\epsilon \int_0^1 \!\ud u  \int_{\mathbb{R}^d}  \left|V(x+z\sigma \sqrt{2\epsilon} )\right| \ud  \mu_{ \sqrt{ u}}(z).
\end{eqnarray*}
Even more, the integral over $\mathbb{R}^d$ in the last relation can be restricted to the ball $\|z\|< 1$.  Indeed if $\|z\|\geq 1$,  we have 
\begin{eqnarray*}
\left({{2 \pi u}} \right)^{-d/2}\exp\left\{{-\frac{\|z\|^2}{2u}}\right\} = {u}^{-d/2} \exp\left\{-\frac{\|z\|^2}{2}\left(\frac{1}{u}-1 \right)\right\} \\\times \left({{2 \pi}} \right)^{-d/2}\exp\left( -{\|z\|^2}/{2}\right) \leq 
u^{-d/2} e^{1/2}e^{-1/(2u)}\\\times\left({{2 \pi}} \right)^{-d/2}\exp\left( -{\|z\|^2}/{2}\right).\nonumber 
\end{eqnarray*}
Notice that \[M_0=\sup_{u>0}e^{1/2}u^{-d/2}\exp[-1/(2u)]<\infty.\] Then, a little calculus and  Th.~\ref{th:1A}.c) show that
\begin{eqnarray*}&&
\epsilon \int_0^1 \!\ud u  \int_{\|z\|\geq1} \left|V(x+z\sigma \sqrt{2\epsilon} )\right| \ud  \mu_{ \sqrt{ u}}(z)\\&& \leq \epsilon M_0 \int_{\mathbb{R}^d}  \left|V(x+z\sigma \sqrt{2\epsilon} )\right|  \ud \mu_{1}(z) < \infty.
\end{eqnarray*} 

To conclude the theorem, we only need to prove that 
\begin{eqnarray*}
\epsilon \int_0^1 \!\ud u  \int_{\|z\|<1} \left|V(x+z\sigma \sqrt{2\epsilon} )\right| \ud  \mu_{ \sqrt{ u}}(z) < \infty
\end{eqnarray*} 
for $\epsilon$ small enough. Pick an arbitrary $\eta >0$.  Remembering that $\epsilon <1/2$ and taking $x$ such that $\|x\|<\eta$,  we notice that in order to compute the integral over $z$ in the above formula, we only need to know the potential over the ball $D$ of radius $\eta+\max_{1\leq i\leq d} \sigma_i$ centered at origin. Therefore, if we set $V_D=1_D V$, then 
\begin{eqnarray}
\label{eq:6A}\nonumber 
\epsilon \int_0^1 \!\ud u  \int_{\|z\|<1} \left|V(x+z\sigma \sqrt{2\epsilon} )\right| \ud  \mu_{ \sqrt{ u}}(z)\\  \leq  \epsilon \int_0^1 \!\ud u  \int_{\mathbb{R}^d} \left|V_D(x+z\sigma \sqrt{2\epsilon} )\right| \ud  \mu_{ \sqrt{ u}}(z)
\end{eqnarray} 
for all $x$ such that $\|x\|< \eta$. Since $D$ is bounded and $V\in K_d^{\text{loc}}$, it follows that $V_D \in K_d$ and then Eq.~(\ref{eq:3b}) guaranties that there is $\epsilon_0>0$ such that the last integral in Eq.~(\ref{eq:6A}) is uniformly bounded for all $x$. Consequently,   
\begin{eqnarray*}
\epsilon_0 \int_0^1 \!\ud u  \int_{\|z\|<1} \left|V(x+z\sigma \sqrt{2\epsilon_0} )\right| \ud  \mu_{ \sqrt{ u}}(z) 
\end{eqnarray*} 
is bounded for all $x$ such that $\|x\|< \eta$. Since $\eta $ is arbitrary, we are done. 
 $\quad \Box$
 
 \begin{6}
\label{th:3A}
Assume $V$ is a Kato-class potential. Then there is $M_\beta >0$ a constant depending upon the inverse temperature $\beta$ such that $\rho(x,x';\beta) \leq M_\beta $ for all $(x,x')\in \mathbb{R}^d \times \mathbb{R}^d$. 
\end{6}

\emph{Observation.} The proof of this theorem will show why the Kato class is the natural class for the treatment of Feynman-Ka\c{c} semigroups. Most of the arguments used in the proof are borrowed from Aizenman and Simon.\cite{Aiz82} 

\emph{Proof of the theorem.}  
If $V_-$ denotes the negative part of $V$, we  notice that
\begin{eqnarray*}&& 
\mathbb{E} \exp\left\{-\beta \int_0^1 V[x_0(u)+\sigma B_u^0]\right\}\\&& \leq \mathbb{E} \exp\left\{\beta \int_0^1 V_-[x_0(u)+\sigma B_u^0]\right\}
\end{eqnarray*}
so, without loss of generality, we may assume that $V$ is of class $K_d$. The proof of this theorem is organized in three steps, each step reducing the problem to a simpler statement. 

\emph{Step~1.} In the first step, we prove that it suffices to show that 
\begin{equation}
\label{eq:7A}
\sup_{x} \int_{\mathbb{R}^d} \rho(x,x';\beta)\ud x' < \infty
\end{equation}
for all $\beta > 0$ and $ V\in K_d$. 

In this part of the proof, it is convenient to denote the density matrix by $\rho_{V}(x,x';\beta)$, the index $V$ indicating the potential from which the density matrix is derived. For the proof, we need two well-known properties of the density matrix $\rho_{V}(x,x';\beta)$: it is symmetrical 
\[
\rho_V(x,x';\beta)=\rho_V(x',x;\beta)
\]
and it satisfies the Trotter product rule
\[
\rho_V(x,x';\beta)=\int_{\mathbb{R}^d} \rho_V(x,y;\beta/2)\rho_V(y,x';\beta/2) \ud y.
\] 
These two properties can be established by direct computation starting with the definition of the Brownian bridge. The first one is a consequence of the symmetry of the standard Brownian bridge, that is $\{B_{1-u}^0: 0\leq u \leq 1\}$ is also a Brownian bridge and is equal in distribution to $\{B_{u}^0: 0\leq u \leq 1\}$. The Trotter product rule is a consequence of the Markov property of the Brownian motion $B_u$ entering the definition of the Brownian bridge. The simple proofs are left to the reader. 

Now, the Cauchy-Schwartz inequality gives the estimate
\begin{eqnarray*}
\rho_V(x,x';\beta)\leq\left[\int_{\mathbb{R}^d} \rho_V(x,y;\beta/2)^2 \ud y \right]^{1/2} \\ \times \left[\int_{\mathbb{R}^d} \rho_V(y,x';\beta/2)^2 \ud y \right]^{1/2}.
\end{eqnarray*}
Taking the supremum over $x$ and $x'$ and using the symmetry of the density matrix, one concludes that 
\begin{equation}
\label{eq:8A}
\sup_{x,x'} \rho_V(x,x';\beta)\leq \sup_{x} \int_{\mathbb{R}^d} \rho_V(x,y;\beta/2)^2 \ud y. 
\end{equation} 
Again by the Cauchy-Schwartz inequality, 
\begin{eqnarray*}
\frac{\rho_V(x,x';\beta/2)^2}{\rho_{fp}(x,x';\beta/2)^2}\\ = \left(\mathbb{E}\exp\left\{-\frac{\beta}{2} \int_0^1V\left[x_0(u)+\frac{\sigma}{\sqrt{2}} B_u^0\right]\ud u\right\}\right)^2\\
\leq \mathbb{E}\exp\left\{-\frac{\beta}{2} \int_0^1 2V\left[x_0(u)+\frac{\sigma}{\sqrt{2}} B_u^0\right]\ud u\right\}.
\end{eqnarray*}
Next, we combine the last equation with the bound
\[
\rho_{fp}(x,x';\beta/2)^2\leq \left(\prod_{i=1}^d\frac{1}{\sqrt{\pi \sigma^2_i}}\right)\rho_{fp}(x,x';\beta/2),
\]
to obtain the inequality
\begin{equation}
\label{eq:9A}
\rho_{V}(x,x';\beta/2)^2\leq \left(\prod_{i=1}^d\frac{1}{\sqrt{\pi \sigma^2_i}}\right) {\rho_{2V}}(x,x';\beta/2).
\end{equation}
Substituting Eq.~(\ref{eq:9A}) in Eq.~(\ref{eq:8A}), one obtains
\begin{equation*}
\sup_{x,x'} \rho_V(x,x';\beta)\leq \left(\prod_{i=1}^d\frac{1}{\sqrt{\pi \sigma^2_i}}\right)\sup_{x} \int_{\mathbb{R}^d} \rho_{2V}(x,y;\beta/2) \ud y
\end{equation*} 
and the claim of Step~1 is concluded because $\beta/2 >0$ and $2V\in K_d$. 

\emph{Step~2.} Simple transformations of coordinates show that 
\begin{eqnarray*}
\int_{\mathbb{R}^d} \rho(x,x';\beta)\ud x'=\int_{\mathbb{R}^d} \ud \mu_1(z) \mathbb{E}\, e^{-{\beta} \int_0^1V\left[x+z u \sigma+{\sigma} B_u^0\right]\ud u}
\end{eqnarray*}
and from the very definition of the Brownian bridge, we learn that $z u + B_u^0$ is in fact a Brownian motion $B_u$ starting at zero. Thus, 
\begin{eqnarray*}
\int_{\mathbb{R}^d} \rho(x,x';\beta)\ud x'= \mathbb{E}\, e^{-{\beta} \int_0^1V\left(x+\sigma B_u\right)\ud u}.
\end{eqnarray*}
For the remainder of the proof, $\mathbb{E}$ stands for the expectation value with respect to the underlying measure of the standard Brownian motion $B_u$.

In this second step, we use the Markov property of the Brownian motion to show that if there is $\epsilon_0>0$ such that the inequality
\begin{eqnarray*}
\sup_x \mathbb{E}\, e^{-{\beta} \int_0^\epsilon V\left(x+\sigma B_u\right)\ud u} < \infty
\end{eqnarray*}
holds for all $\epsilon < \epsilon_0$, then Eq.~(\ref{eq:7A}) also holds.

Let $\theta $ and $\tau $ be some positive real numbers such that $\theta+ \tau=1$. We  break the integrand in the above equation in two parts
\begin{eqnarray*}
e^{-{\beta} \int_0^1V\left(x+\sigma B_u\right)\ud u}=e^{-{\beta} \int_0^\theta V\left(x+\sigma B_u\right)\ud u} \\ \times e^{-{\beta} \int_\theta^1V\left(x+\sigma B_u\right)\ud u}=e^{-{\beta} \int_0^\theta V\left(x+\sigma B_u\right)\ud u} \\ \times e^{-{\beta} \int_0^\tau V\left(x+\sigma B_{\theta+u}\right)\ud u}. 
\end{eqnarray*}
Using the Markov property, we learn that the expected value of the above integrand conditioned on the random variables $B_u$ with $u\in [0,\theta]$ is
\begin{eqnarray*}
e^{-{\beta} \int_0^\theta V\left(x+\sigma B_u\right)\ud u} \mathbb{E}'  e^{-{\beta} \int_0^\tau V\left(x+\sigma B_{\theta}+\sigma B'_{u}\right)\ud u}. 
\end{eqnarray*}
Here, the symbol $\mathbb{E}'$ denotes the expected value against the standard Brownian motion starting at zero $B'_u$, Brownian motion that is independent from $B_u$. 
The above conditional expectation is smaller or equal than 
\begin{eqnarray*}
e^{-{\beta} \int_0^\theta V\left(x+\sigma B_u\right)\ud u} \sup_{x}\mathbb{E}\,  e^{-{\beta} \int_0^\tau V\left(x+\sigma B_{u}\right)\ud u},
\end{eqnarray*}
where the prime sign becomes superfluous and is therefore dropped. 
Taking the total expectation and then the supremum over $x$, we obtain the inequality
\begin{eqnarray*}
\sup_{x}\mathbb{E}\, e^{-{\beta} \int_0^1 V\left(x+\sigma B_{u}\right)\ud u} \leq  
\sup_{x} \mathbb{E}\, e^{-{\beta} \int_0^\theta V\left(x+\sigma B_u\right)\ud u} \\ \times  \sup_{x}\mathbb{E}\,  e^{-{\beta} \int_0^\tau V\left(x+\sigma B_{u}\right)\ud u}. 
\end{eqnarray*}
A simple inductive argument then shows that
\begin{equation}
\label{eq:10A}
\sup_{x}\mathbb{E} \, e^{-{\beta} \int_0^1 V\left(x+\sigma B_{u}\right)\ud u}\leq  
\left\{\sup_{x} \mathbb{E}\, e^{-{\beta} \int_0^{1/n} V\left(x+{\sigma} B_u\right)\ud u} \right\}^n
\end{equation}
for all $n \geq 1$. Clearly, the claim of Step~2 is concluded because the right-hand side of Eq.~(\ref{eq:10A}) is finite for all  $n$ such that $1/n < \epsilon_0$.  

\emph{Step~3}. In this final step, we prove that there is $\epsilon_0$ small enough such that 
\begin{eqnarray*}
 \sup_x \mathbb{E}\, e^{-{\beta} \int_0^\epsilon V\left(x+\sigma B_u\right)\ud u} < \infty 
\end{eqnarray*}
for all $\epsilon < \epsilon_0$. Eq.~(\ref{eq:3a}) allows us to pick some $\epsilon_0>0$ such that 
\begin{equation}
\label{eq:11A}
\sup_x \mathbb{E}\, \left[{\beta} \int_0^\epsilon|V|\left(x+\sigma B_u\right)\ud u\right]< 1/2
\end{equation}
for all $\epsilon < \epsilon_0$. Next, we consider the inequality
\begin{eqnarray}
\label{eq:12A}\nonumber
\sup_x \mathbb{E}\, e^{-{\beta} \int_0^\epsilon V\left(x+\sigma B_u\right)\ud u}\leq  \sup_x \mathbb{E}\, e^{{\beta} \int_0^\epsilon |V|\left(x+\sigma B_u\right)\ud u}\\=  \sup_x \sum_{k=1}^{\infty}  \frac{\beta^k}{k!}\mathbb{E} \left[\int_0^\epsilon|V|\left(x+\sigma B_u\right)\ud u\right]^k \leq \sum_{k=0}^\infty A_k, \quad
\end{eqnarray}
where 
\begin{eqnarray}
\label{eq:13A}\nonumber
A_k= \sup_x  \frac{\beta^k}{k!}\mathbb{E} \left[\int_0^\epsilon|V|\left(x+\sigma B_u\right)\ud u\right]^k \\=
  {\beta^k}\mathbb{E} \int_{0\leq s_1 \leq \ldots \leq s_{k}\leq \epsilon} |V|\left(x+\sigma B_{s_1}\right)\\ \nonumber \ldots |V|\left(x+\sigma B_{s_k}\right)\ud s_1 \ldots \ud s_k.
\end{eqnarray}
Notice that the  term by term integration of the first series appearing in Eq.~(\ref{eq:12A}) is guaranteed by the Monotone convergence theorem. The last equality in Eq.~(\ref{eq:13A}) follows by symmetry arguments. 

To construct a bound for the terms $A_k$, we first condition  on the random variables $B_u$ with $u\in [0,s_{k-1}]$ and use  the Markov property of $B_u$ to show that this conditional expectation has the value
\begin{eqnarray}
\label{eq:14A}\nonumber
{\beta^{k-1}}\int_{0\leq s_1 \leq \ldots \leq s_{k-1}\leq \epsilon} |V|\left(x+\sigma B_{s_1}\right) \ldots |V|\left(x+\sigma B_{s_{k-1}}\right)\\ \times \left[ \mathbb{E}' \beta \int_{0}^{\epsilon-s_{k-1}}  |V|\left(x+\sigma B_{s_{k-1}}+\sigma B'_{u}\right)\ud u\right]\ud s_1 \ldots \ud s_{k-1}.\quad
\end{eqnarray}
Eq.~(\ref{eq:11A}) shows that the quantity in the square brackets is bounded by $1/2$. Therefore, the conditional expectation given by Eq.~(\ref{eq:14A}) is bounded by
\begin{eqnarray}
\label{eq:15A} \nonumber \frac{1}{2}
 {\beta^{k-1}}\int_{0\leq s_1 \leq \ldots \leq s_{k-1}\leq \epsilon} |V|\left(x+\sigma B_{s_1}\right) \\ \ldots |V|\left(x+\sigma B_{s_{k-1}}\right) \ud s_1 \ldots \ud s_{k-1}.
\end{eqnarray}
 Taking the total expectation in the Eqs.~(\ref{eq:14A}) and (\ref{eq:15A})  and then the supremum over $x$, we learn that \[A_k \leq A_{k-1}/2,\] from which the inequality $A_k \leq 1/2^k$ follows by induction. Substituting  this last inequality in Eq.~(\ref{eq:12A}), we obtain
\begin{eqnarray*}
 \sup_x \mathbb{E}\, e^{-{\beta} \int_0^\epsilon V\left(x+\sigma B_u\right)\ud u} \leq \sum_{k=0}^\infty \frac{1}{2^k} =2 < \infty
\end{eqnarray*}
for all $\epsilon < \epsilon_0$ and the proof of Step~3 and of the theorem is concluded. $\quad \Box$


\begin{thebibliography}{99}
\bibitem{Fey48} R. P. Feynman,  Rev. Modern Phys. \textbf{20}, 367 (1948).
\bibitem{Kac51} M. Ka\c{c}, in Proceedings of the 2nd Berkeley Symposium on Mathematical Statistics and Probability (University of California, Berkeley, 1951) pp.~189-215.
\bibitem{Dol99r} J. D. Doll, M. Eleftheriou, S. A. Corcelli, and David L. Freeman, \emph{Quantum Monte Carlo Methods in Physics and Chemistry,} edited by M.P. Nightingale and C.J. Umrigar, NATO ASI Series, Series C Mathematical and Physical Sciences, Vol. X, (Kluwer, Dordrecht, 1999).
\bibitem{Met53} N. Metropolis, A. W. Rosenbluth, M. N. Rosenbluth, A. M. Teller and E. Teller, J. Chem. Phys. \textbf{21}, 1087 (1953).
\bibitem{Kal86} M. Kalos and P. Whitlock, \emph{Monte Carlo Methods} (Wiley-Interscience, New York, 1986).
\bibitem{Mie01} S. L. Mielke and  D. G. Truhlar, J. Chem. Phys. \textbf{114}, 621 (2001).
\bibitem{Pre02} C. Predescu and J. D. Doll, J. Chem. Phys. \textbf{117}, 7448 (2002).
\bibitem{Dol85} J. D. Doll, R. D. Coalson, and D. L. Freeman, Phys. Rev. Lett. \textbf{55}, 1 (1985).
\bibitem{Dol84} J. D. Doll and D. L. Freeman, J. Chem. Phys. \textbf{80}, 2239 (1984).
\bibitem{Ele99} M. Eleftheriou, J. D. Doll, E. Curotto, and D. L. Freeman, J. Chem. Phys.
\textbf{110}, 6657 (1999).
\bibitem{Law69} S. V. Lawande, C. A. Jensen, and H. L. Sahlin, J. Comput. Phys. \textbf{3}, 416
(1969).
\bibitem{Kle95} H. Kleinert, \emph{Path Integrals in Quantum Mechanics, Statistics
and Polymer Physics} (World Scientific, Singapore, 1995).
\bibitem{Ale90} C. Alexandrou, W. Fleischer, and R. Rosenfelder, Phys. Rev. Lett. \textbf{65}, 2615 (1990); Phys. Rep. \textbf{215}, 1 (1992). 
\bibitem{Tit01} J. T. Titantah, C. Pierleoni, and S. Ciuchi,  Phys. Rev. Lett. \textbf{87} 206404-1 (2001). 
\bibitem{Kol01} J. S. Kole and H. De Raedt, Phys. Rev. E \textbf{64}, 016704 (2001).
\bibitem{Mus97} M. H. M\"user and B. J. Berne, J. Chem. Phys. \textbf{107}, 571 (1997).
\bibitem{Sim79} B. Simon, \emph{Functional Integration and Quantum Physics} (Academic, London, 1979).
\bibitem{Sim82} B. Simon, Bull. AMS, \textbf{7}, 447-526 (1982).
\bibitem{Aiz82} M. Aizenman and B. Simon, Commun. Pure  Appl. Math \textbf{35}, 209 (1982).
\bibitem{Kwa92} S. Kwapien and W. A. Woyczynski, \emph{Random Series and Stochastic Integrals: Single and Multiple} (Birkh\"auser, Boston, 1992), Theorem~2.5.1.
\bibitem{Dur96} R. Durrett, \emph{Probability: Theory and Examples,} 2nd ed. (Duxbury, New York, 1996).
\end{thebibliography}
\end{document}